\documentclass[journal]{IEEEtran}
\usepackage{amssymb,latexsym,graphicx,times,amsmath,subfigure,threeparttable,booktabs,bm,color,multirow,amsmath,amsfonts,amsthm}
\newcommand{\tabincell}[2]{\begin{tabular}{@{}#1@{}}#2\end{tabular}}

\begin{document}

\title{Revisiting Euclidean Alignment for Transfer Learning in EEG-Based Brain-Computer Interfaces}

\author{Dongrui~Wu

\thanks{D. Wu is with the Ministry of Education Key Laboratory of Image Processing and Intelligent Control, Huazhong University of Science and Technology, Wuhan, 430074 China. He is also  with Zhongguancun Academy, Beijing, 100084 China. Email: drwu09@gmail.com.}}

\maketitle

\begin{abstract}
Due to large intra-subject and inter-subject variabilities of electroencephalogram (EEG) signals, EEG-based brain-computer interfaces (BCIs) usually need subject-specific calibration to tailor the decoding algorithm for each new subject, which is time-consuming and user-unfriendly, hindering their real-world applications. Transfer learning (TL) has been extensively used to expedite the calibration, by making use of EEG data from other subjects/sessions. An important consideration in TL for EEG-based BCIs is to reduce the data distribution discrepancies among different subjects/sessions, to avoid negative transfer. Euclidean alignment (EA) was proposed in 2020 to address this challenge. Numerous experiments from 13 different BCI paradigms demonstrated its effectiveness and efficiency. This paper revisits EA, explaining its procedure and correct usage, introducing its applications and extensions, and pointing out potential new research directions. It should be very helpful to BCI researchers, especially those who are working on EEG signal decoding.
\end{abstract}


\IEEEpeerreviewmaketitle

\section{Introduction}

Brain-computer interfaces (BCIs) enable direct communication between the brain and external devices, which can be used in text input \cite{Nakanishi2020}, device control \cite{Leeb2007}, neuro-rehabilitation \cite{Mane2020}, emotion recognition and regularization \cite{drwuPIEEE2023}, speech decoding \cite{Moses2021}, restoring walking abilities \cite{Lorach2023}, etc. It is an important component of Brain Initiatives/Projects all over the world, particularly China \cite{Liu2023} and the IEEE\footnote{https://brain.ieee.org/}. It was selected by Nature as one of seven technologies to watch in 2024 \cite{Eisenstein2024}, and again one of eight technology events to watch for in 2025 \cite{Naddaf2025}, and also selected by IEEE Computer Society in 2025 as one of ``\emph{22 breakthrough technologies set to redefine industries and shape the future of our world for decades to come}" \cite{Milojicic2025}.

According to the input signals, BCIs can be broadly classified into invasive ones and non-invasive ones. The former use surgery to implant electrodes into the brain, which are mainly used for patients. Non-invasive BCIs do not need surgery, and hence are more convenient and preferred by able-bodied users. Electroencephalogram (EEG) is the most frequently used input signal for non-invasive BCIs.

EEG signals exhibit both intra-subject variability (non-stationarity), e.g., the same subject may have different EEG responses to the same stimulus in different sessions, and inter-subject variability (individual differences), i.e., different subjects may have significantly different EEG responses to the same stimulus. As a result, it is very challenging, if not impossible, to design an EEG-based non-invasive BCI system that is plug-and-play and fits all subjects. Usually, for each new subject, we need to collect some subject-specific calibration data to tailor the parameters of the decoding algorithm, which is time-consuming and user-unfriendly.

Transfer learning (TL), which utilizes data/knowledge from other subjects (sessions) to facilitate the calibration for a new subject (session), has been extensively used in EEG-based BCIs to expedite the calibration \cite{drwuTLBCI2022}. An important consideration in TL is how to reduce the EEG data distribution discrepancies among different subjects (sessions) to avoid negative transfer \cite{drwuNTL2023}.

Euclidean Alignment (EA) \cite{drwuEA2020} was proposed in 2020 to address this challenge. It uses two simple, efficient and closed-form formulas to align the raw EEG trials from different subjects/sessions, greatly facilitating TL \cite{drwuMITLBCI2022}. It also inspires some other EEG data alignment approaches, e.g., label alignment (LA) \cite{drwuLA2020}. The EA paper has been cited 283 times in WoS and 405 times in Google Scholar (as of May 3, 2025), ranking 2/344 of all papers published in IEEE Transactions on Biomedical Engineering\footnote{https://webofscience.clarivate.cn/wos/alldb/summary/fd21b6a6-bdf1-4dcd-bae6-45d4041b195f-014738bfd9/times-cited-descending/1} (TMBE) in 2020, and 3/1872 of all papers published in IEEE TBME\footnote{https://webofscience.clarivate.cn/wos/alldb/summary/5e219ca8-1532-4637-a27d-6c70f9cbbca2-014738a148/times-cited-descending/1} between 2020 and 2025. It was one of the only two Highly Cited Papers of IEEE TBME in 2020. It was also the number one contributor to the 2022 impact factor (4.6) of IEEE TBME\footnote{https://jcr.clarivate.com/jcr-jp/journal-profile?journal=IEEE\%20T\%20BIO-MED\%20ENG\&year=2022\&fromPage=\%2Fjcr\%2Fhome}, among all 695 citable items published in 2020 and 2021.

After five years of the publication of the EA paper, we now have deeper and more comprehensive understanding on how to properly use EA. Its applicable BCI paradigms have also been greatly expanded. This paper revisits the EA approach, explaining its procedure and correct usage, introducing its applications and extension, and pointing out potential new research directions. It should be very helpful to BCI researchers, especially those who are working on EEG signal decoding.

The remainder of this paper is organized as follows: Section~\ref{sect:EA} introduces the procedure, correct usage, and applications of EA. Section~\ref{sect:ext} introduces an extension of EA, and some related EEG data alignment approaches. Section~\ref{sect:future} points out potential future research directions. Finally, Section~\ref{sect:conclusions} draws conclusions.

\section{Euclidean Alignment (EA)} \label{sect:EA}

This section introduces the EA approach \cite{drwuEA2020}. It was inspired by Riemannian alignment \cite{Zanini2018} proposed in 2018 in IEEE TBME, which has been cited 272 times in WoS\footnote{https://webofscience.clarivate.cn/wos/alldb/summary/bbb3db2d-ab56-4291-8b2e-884a88511745-015ed46b9e/times-cited-descending/1} and 386 times in Google Scholar (as of may 3, 2025) and is one of the five Highly Cited Papers of IEEE TBME in 2018. Riemannian alignment has demonstrated promising performance in seven different BCI paradigms, as shown in Table~\ref{tab:RA}.

\begin{table*}[htbp]\centering \setlength{\tabcolsep}{2mm}
\caption{Classification accuracy improved by Riemannian alignment in 7 different BCI paradigms. Note that different publications generally used different datasets.} \label{tab:RA}
\begin{tabular}{c|c|c|c|c|c|c}
\toprule
\multirow{2.5}{*}{No.}&\multirow{2.5}{*}{\tabincell{c}{BCI\\ Paradigm}} & \multirow{2.5}{*}{Classifier}& \multicolumn{3}{c|}{Accuracy (\%)} & \multirow{2.5}{*}{Reference}\\ \cmidrule(lr){4-6}
&& & w/o RA & w/ RA& Improvement & \\ \midrule
\multirow{30}{*}{1} &\multirow{30}{*}{\tabincell{c}{Motor\\ Imagery}} & MDM & 38.84 & 63.76 &\textbf{24.92} & \multirow{2}{*}{Zanini et al., IEEE TBME 2018  \cite{Zanini2018}} \\
                    && GM-4 & 41.58 & 64.18 & \textbf{22.60}  &  \\ \cline{7-7}
                    &&  MDM & 50.14 & 66.21 & { \textbf{16.07}} & He and Wu, IEEE SMC 2019\cite{drwuSMC2019} \\ \cline{7-7}
                    && MDM & 64.81 & 66.82 & { \textbf{2.01}} & \multirow{2}{*}{Kumar et al., BCI 2019 \cite{Kumar2019BCI}} \\
                    && FgMDM & 68.94 & 71.10 & { \textbf{2.16}} &  \\ \cline{7-7}
                    && MDM & 57.18 & 70.91 & { \textbf{13.73}} & He and Wu, IEEE TBME 2020 \cite{drwuEA2020} \\ \cline{7-7}
                    && FBCSP-LDA & 69.00 & 70.00 & { \textbf{1.00}} & \multirow{3}{*}{Xu et al., FHN 2020 \cite{Xu2020}} \\
                    && ShallowNet & 71.00 & 72.00 & { \textbf{1.00}} &  \\
                    && EEGNet & 70.00 & 72.00 & { \textbf{2.00}} &  \\ \cline{7-7}
                    && EEGNet & 69.70 & 71.40 & { \textbf{1.70}} & \multirow{2}{*}{Xu et al., JNE 2021 \cite{Xu2021JNE}} \\
                    && EEGNetv4 & 67.40 & 70.30 & { \textbf{2.90}} & \\ \cline{7-7}
                    && MDM & 57.02 & 72.45 & { \textbf{15.43}} & Zhang et al., CMPB 2021 \cite{Zhang2021CMPB} \\
                    && MDM & 57.18 & 70.91 & { \textbf{13.73}} & Liang et al., IEEE IROS 2021 \cite{Liang2021IROS} \\
                    && MDM & 56.00 & 70.00 & { \textbf{14.00}} & Li et al., IEEE Access 2021 \cite{Li2021} \\
                    && MDM & 57.18 & 70.91 & { \textbf{13.73}} & Liang et al., ICIEA 2022 \cite{Liang2022} \\
                    && EEGNet & 53.45 & 57.93 & { \textbf{4.48}} & Zoumpourlis and Patras, BCI 2022 \cite{Zoumpourlis2022} \\
                    && MDM & 32.75 & 50.93 & { \textbf{18.18}} & Xia et al., IEEE TBME 2022 \cite{drwuTBME2022} \\
                    && MDM & 75.70 & 76.73 & { \textbf{1.03}} & Pan et al., JNE 2023 \cite{Pan2023JNE} \\
                    && MDM & 57.18 & 70.91 & { \textbf{13.73}} & Liang et al., BSPC 2024 \cite{Liang2024BSPC} \\
                    && BaseNet & 57.38 & 63.64 & { \textbf{6.26}} & Wimpff et al., BCI 2024 \cite{Wimpff2024} \\ \cline{7-7}
                    && EEGNet & 72.18 & 76.42 & { \textbf{4.24}} &  \multirow{3}{*}{Junqueira et al., JNE 2024 \cite{Junqueira2024}} \\
                    && DeepNet & 69.27 & 76.92 & { \textbf{7.65}} &  \\
                    && ShallowNet & 69.99 & 74.23 & { \textbf{4.24}} &   \\ \cline{7-7}
                    && MDM & 57.18 & 70.91 & { \textbf{13.73}} & Chu et al., JNE 2024 \cite{Chu2024} \\ \cline{7-7}
                    && MDM & 54.36 & 73.57 & { \textbf{19.21}} & \multirow{6}{*}{Jiang et al., ICASSP 2024 \cite{Jiang2024a}} \\
                    && LDA & 52.07 & 70.64 & { \textbf{18.57}} &  \\
                    && CSP-MDM & 61.78 & 87.93 & { \textbf{26.15}} & \\
                    && CSP-LDA & 65.71 & 83.64 & { \textbf{17.93}} &  \\
                    && JF-MDM & 82.36 & 94.3 & { \textbf{11.94}} & \\
                    && JF-LDA & 80.21 & 94.36 & { \textbf{14.15}} &  \\ \midrule
\multirow{2}{*}{2} &\multirow{2}{*}{\tabincell{c}{Event Related\\ Potential}}   & MDM & 51.43 & 87.48 &\textbf{36.05} & \multirow{2}{*}{Zanini et al., IEEE TBME 2018\cite{Zanini2018}} \\
                    && GM-4 & 45.99 & 86.18 & \textbf{40.19}  &\\ \midrule
 \multirow{2}{*}{3} &\multirow{2}{*}{Mental Imagery} & MDM & 58.33 & 69.71 & { \textbf{11.38}} & \multirow{2}{*}{Kumar et al., BCI 2019 \cite{Kumar2019BCI}} \\
                                 && FgMDM & 58.50 & 70.35 & { \textbf{11.85}} &  \\ \midrule
 4&\tabincell{c}{Steady-State\\ Visual Evoked \\ Potential} &MDM	&69.30	&73.00	&\textbf{3.70}& Rodrigues et al., IEEE TBME 2019 \cite{Rodrigues2019} \\ \midrule
 5&\tabincell{c}{Motor+Speech\\ Imagery}  & MDM & 51.32 & 54.82 & { \textbf{3.50}} & Zhan et al., CBM 2022 \cite{Zhan2022} \\ \midrule
 6&\tabincell{c}{Mindfulness\\ Meditation} & EEGNet & 54.88 & 69.93 & { \textbf{15.05}} & Karaiskou et al.,  MetroXRAINE 2023\cite{Karaiskou2023} \\ \midrule
 7&\tabincell{c}{Natural Hand\\ Movement\\ Decoding} & MDM & 53.33 & 62.99 & { \textbf{9.66}} & Xue et al., Measurement 2025 \cite{Xue2025} \\  \bottomrule
\end{tabular}
\end{table*}

\subsection{The EA Algorithm}

The goal of EA is to make the marginal distributions of EEG data from different subjects/sessions more similar, i.e., to reduce the inter-subject/session EEG data discrepancies, facilitating TL. In fact, after EA, often the EEG data from auxiliary subjects/sessions (called source domains in TL) can be combined directly with the limited amount of calibration data from the new subject/session (called target domain in TL) to train a classifier; the performance is usually quite good even without incorporating any other sophisticated TL techniques.

\begin{figure}[htpb]\centering
\includegraphics[width=\linewidth,clip]{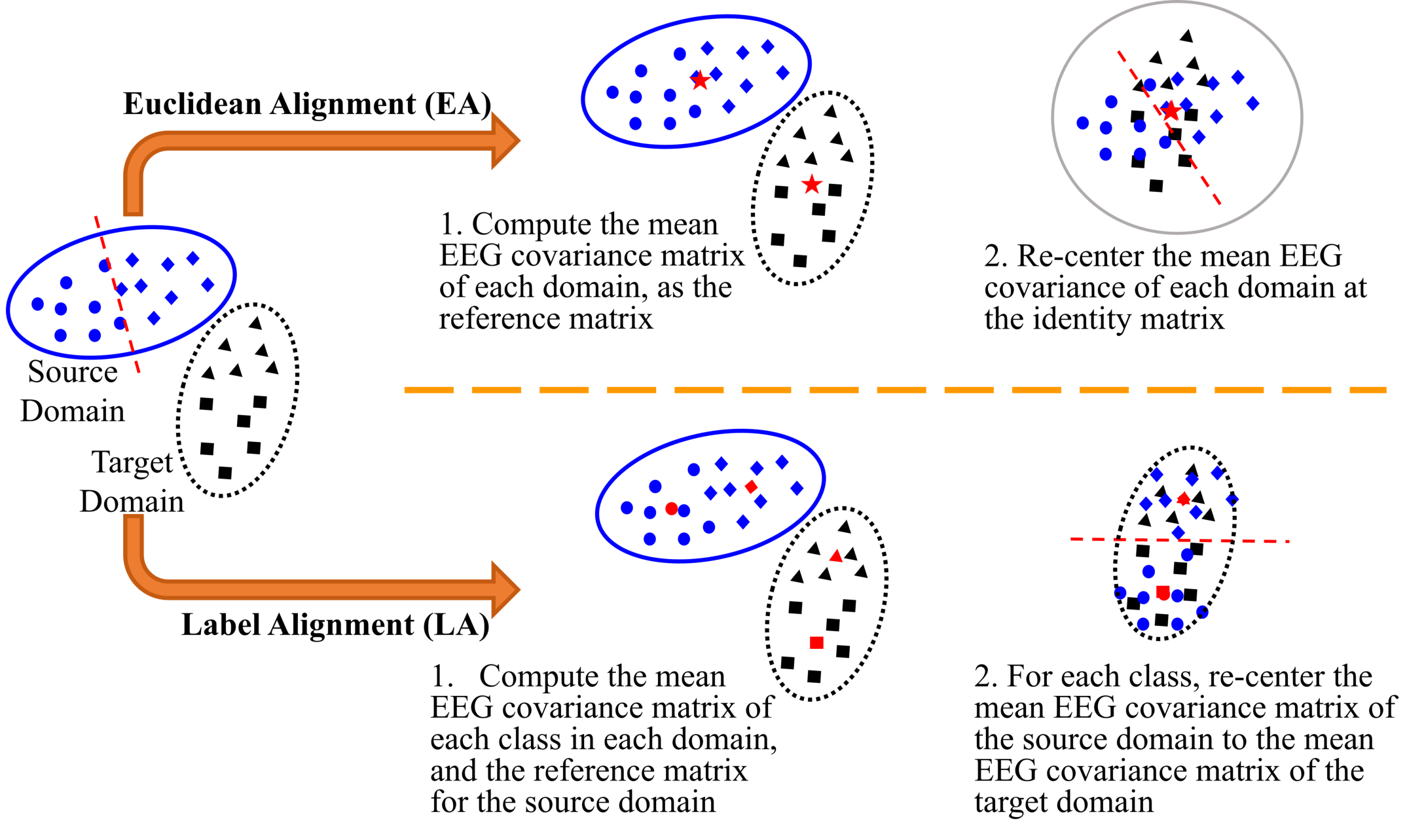}
\caption{Illustration of Euclidean alignment and label alignment. Euclidean alignment aligns the marginal probability distributions of the source and target domains, whereas label alignment aligns the class conditional probability distributions of different domains.} \label{fig:EALA}
\end{figure}

Figure~\ref{fig:EALA} illustrates the flowchart of EA, using only one source domain. Most materials in this subsection are adapted from \cite{drwuEA2020}.

Assume a domain (either source or target; different domains are processed individually and identically) has $N$ EEG trials $\{X_n\}_{n=1}^N$, where $X_n\in \mathbb{R}^{c\times t}$, in which $c$ is the number of EEG channels, and $t$ the number of time-domain sampling points. EA first computes the symmetric and semi positive-definite reference matrix $\bar{R}\in\mathbb{R}^{c\times c}$ as
\begin{align}
  \bar{R}=\frac{1}{N}\sum_{n=1}^N X_nX_n^\top, \label{eq:R}
\end{align}
i.e., $\bar{R}$ is the arithmetic mean of all $N$ covariance matrices from that domain. It then performs the alignment by
\begin{align}
  \tilde{X}_n=\bar{R}^{-1/2}X_n,\quad n=1,...,N. \label{eq:EA}
\end{align}
Here $\bar{R}^{-1/2}$ is called the transformation matrix.

After EA, the mean covariance matrix of all $N$ aligned trials, $\{\tilde{X}_n\}_{n=1}^N$, becomes:
\begin{align}
\frac{1}{N}\sum_{n=1}^N\tilde{X}_n\tilde{X}_n^\top  &=\frac{1}{N}\sum_{n=1}^N\bar{R}^{-1/2}X_nX_n^\top\bar{R}^{-1/2}\nonumber \\
&=\bar{R}^{-1/2}\left(\frac{1}{N}\sum_{n=1}^N X_nX_n^\top\right)\bar{R}^{-1/2}\nonumber\\
&=\bar{R}^{-1/2}\bar{R}\bar{R}^{-1/2}=I, \label{eq:meanC}
\end{align}
i.e., the mean covariance matrix of each domain equals the identity matrix after EA, and hence the EEG data distributions from different domains become more consistent.

EA has the following desirable characteristics \cite{drwuEA2020}:
\begin{enumerate}
\item \emph{Flexible}: It transforms and aligns the EEG trials in the Euclidean space, so any Euclidean space signal processing, feature extraction and machine learning algorithms (most algorithms in the literature belong to the Euclidean space) can then be subsequently applied to them.
\item \emph{Efficient}: It includes only two simple closed-form formulas, which can be computed very fast.
\item \emph{Unsupervised}: It does not need any label information from any domain, so it has much broader applications than supervised algorithms.
\end{enumerate}

\subsection{Where to Place EA in the Traditional BCI TL Pipeline}

\begin{figure}[htpb] \centering
\includegraphics[width=\linewidth,clip]{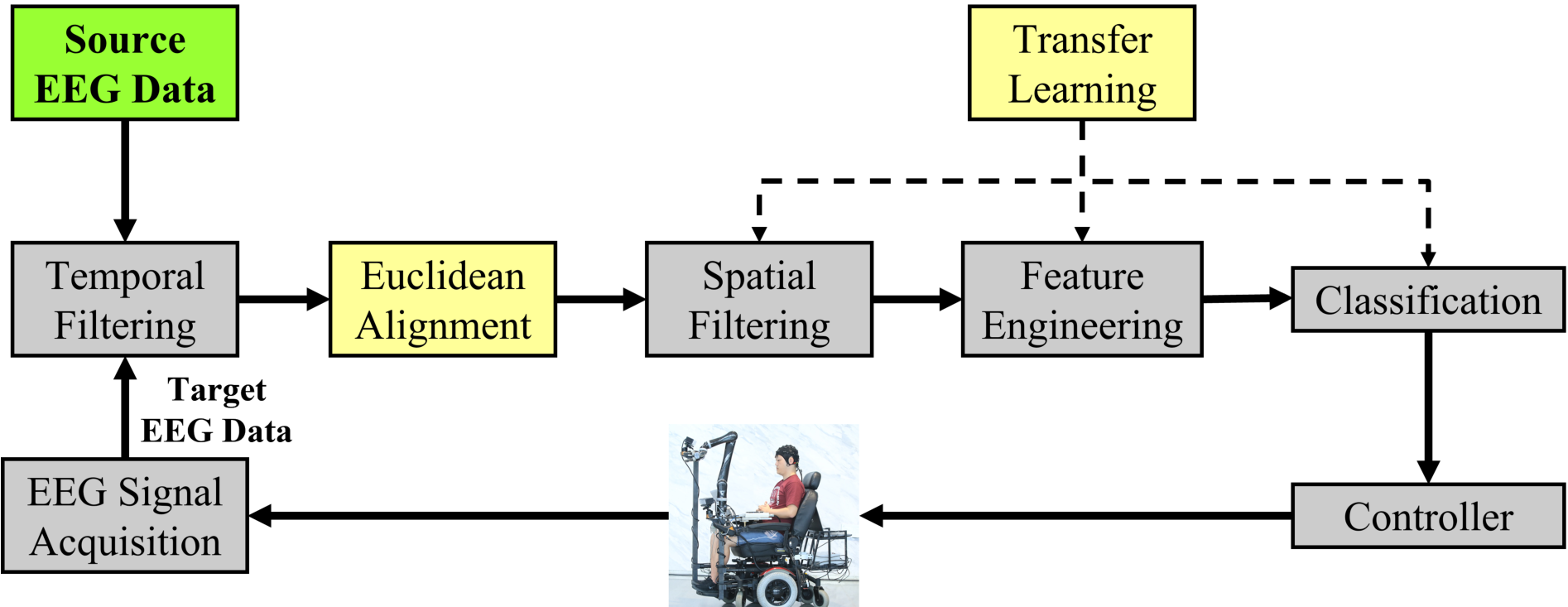}
\caption{The optimal location to perform Euclidean alignment in the traditional BCI transfer learning pipeline (between temporal filtering and spatial filtering).} \label{fig:TL}
\end{figure}

As proposed in our previous research \cite{drwuMITLBCI2022} and illustrated in Figure~\ref{fig:TL}, when traditional machine learning classifiers with manually extracted features are used, EA should be placed between the temporal filtering block and the spatial filtering block. This subsection performs experiments to demonstrate that this placement indeed leads to better performance.

We used the motor imagery Dataset 1 from BCI Competition IV\footnote{http://www.bbci.de/competition/iv/desc\_1.html.} in our experiment, which was also extensively used in our previous research \cite{drwuEA2020,drwuLA2020,drwuMITLBCI2022}. It was recorded from seven healthy subjects using 59 EEG channels at 1000 Hz, and later downsampled to 100 Hz. Each subject performed two classes of motor imagery tasks (100 trials per class) in the calibration session, selected from three options: left hand, right hand, and both feet. Only data from Subjects 2, 3, 4, 5 and 7 were used, as their imagined classes were consistent (they all imagined the movements of left hand and right hand, whereas Subjects 1 and 6 imagined the movements of left hand and both feet). As in \cite{drwuEA2020}, each trial consisted of EEG data between $[0.5, 3.5]$ seconds after the cue appearance, i.e., each trial was a matrix with dimensionality $59\times 300$.

We performed leave-one-subject-out cross-validation in TL: each time, we picked one subject as the test subject (target domain), and combined all trials from the remaining four subjects as the auxiliary data (source domain). The goal was to facilitate the classifier training for the target subject using the source data. The number of labeled calibration trials in the target domain, $M$, increased from 0 to 20, with a step size of 4. The $M$ trials were randomly selected in a continuous block, to avoid the block-design pitfall in BCIs \cite{Li2021b}. The performance of different approaches was evaluated using the classification accuracy on the remaining $100-M$ target domain samples.

We compared three different EA placements, using the best-performing TL pipeline in \cite{drwuMITLBCI2022}:
\begin{enumerate}
\item No EA, as shown in Figure~\ref{fig:TL1}.

\begin{figure}[htpb] \centering
\subfigure[]{\label{fig:TL1} \includegraphics[width=.8\linewidth,clip]{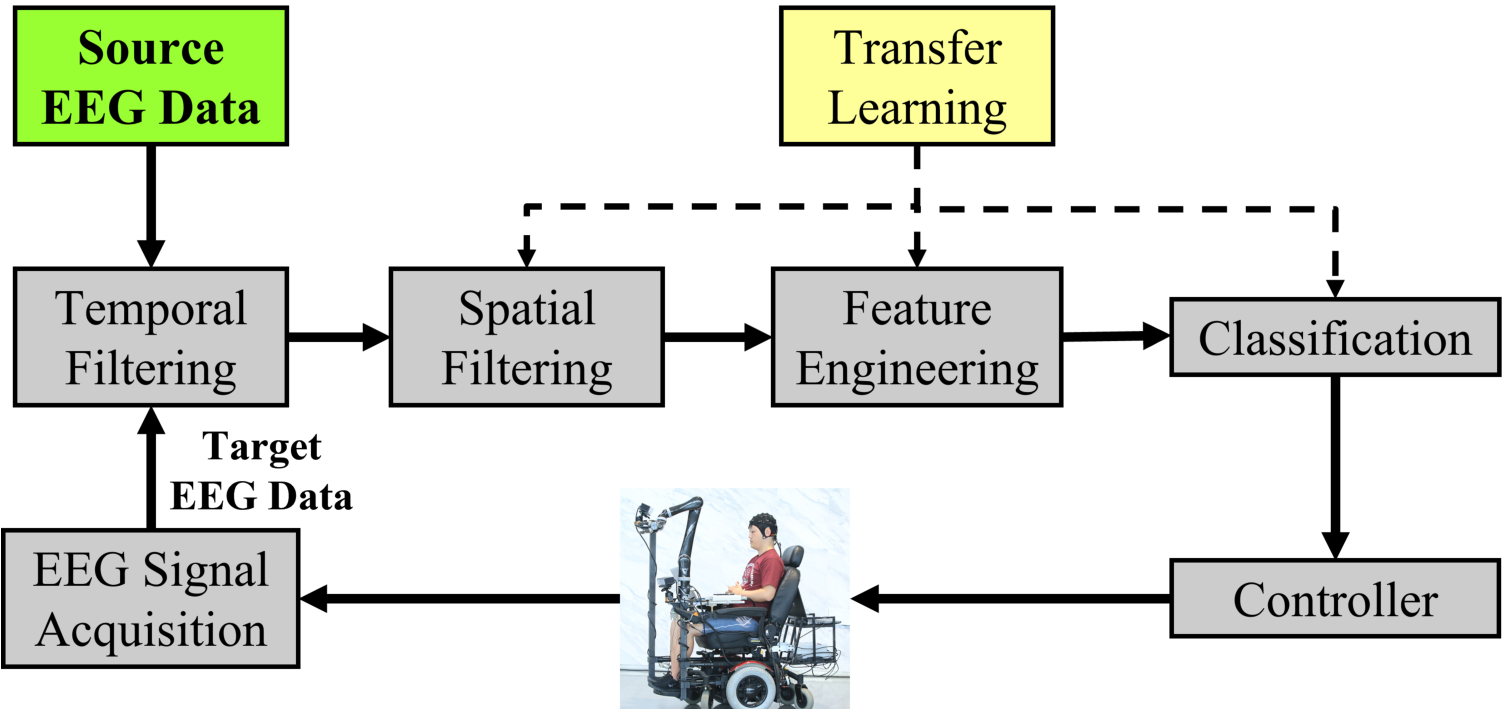}}
\subfigure[]{\label{fig:TL2} \includegraphics[width=\linewidth,clip]{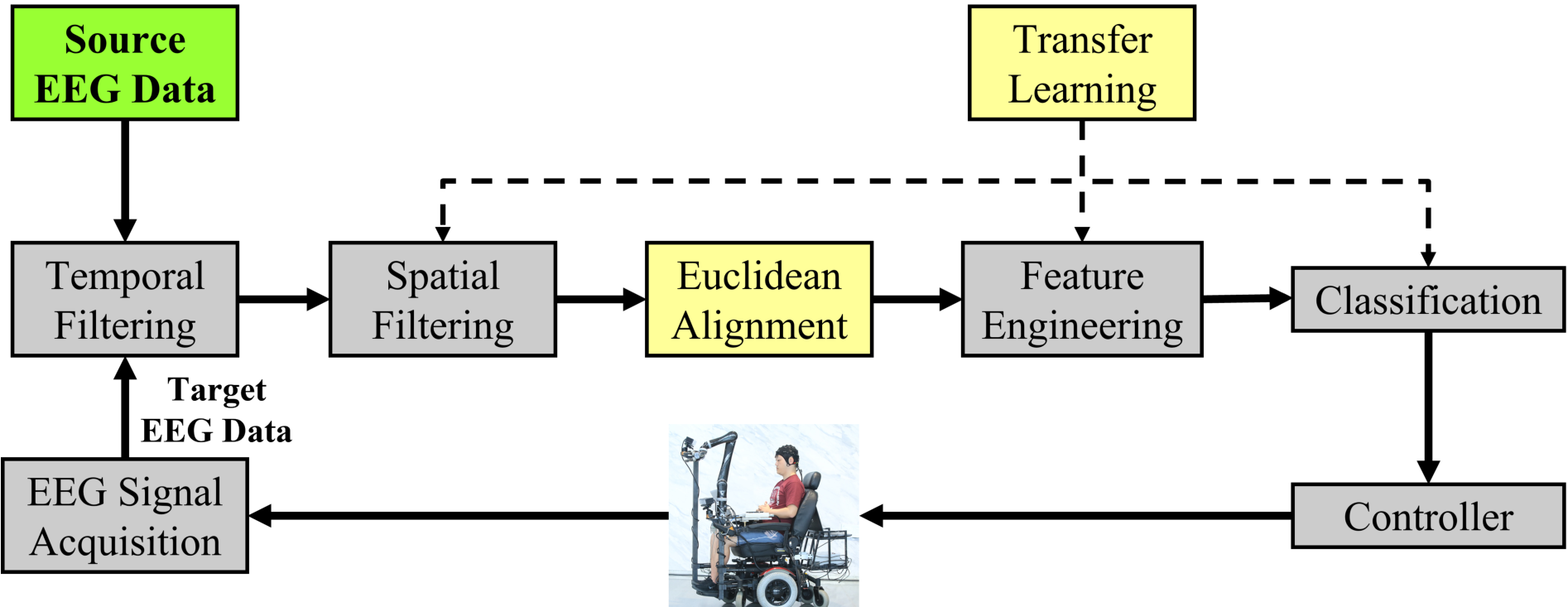}}
\caption{Different configurations of the transfer learning pipeline for motor imagery based BCIs. (a) No Euclidian alignment; (b) performing Euclidean alignment after both temporal filtering and spatial filtering.}
\end{figure}

More specifically, we performed [8,30] Hz band-pass temporal filtering (TF) for each subject's EEGs, epoched them into 3-second trials, combined the (unaligned) trials from all four source subjects as a single source domain, performed regularized common spatial pattern (RCSP) filtering \cite{Lu2010}, extracted the corresponding log-variance features, and finally used weighted adaptation regularization (wAR) \cite{drwuTHMS2017} for classification. This approach is denoted as TF-RCSP-wAR in our analysis.

The above TL pipeline does not re-reference the EEG channels. We also investigated if performing common average referencing (CAR) after temporal filtering can further improve the TL performance. This approach is denoted as TF-CAR-RCSP-wAR.

\item The recommended placement, between temporal filtering and spatial filtering, as shown in Figure~\ref{fig:TL}.

More specifically, we performed [8,30] Hz band-pass temporal filtering for each subject's EEGs, epoched them into 3-second trials, performed EA for each subject individually, combined aligned trials from all four source subjects as a single source domain, performed RCSP filtering \cite{Lu2010}, extracted the corresponding log-variance features, and finally used wAR \cite{drwuTHMS2017} for classification. This approach is denoted as TF-EA-RCSP-wAR in our analysis.

The above TL pipeline did not re-reference the EEG channels. We also evaluated another slightly modified pipeline that performed CAR immediately after temporal filtering, which is denoted as TF-CAR-EA-RCSP-wAR.

\item Placing EA after both temporal filtering and spatial filtering, as shown in Figure~\ref{fig:TL2}.

More specifically, we performed [8,30] Hz band-pass temporal filtering for each subject's EEGs, epoched them into 3-second trials, combined (unaligned) trials from all four source subjects as a single source domain, performed RCSP filtering \cite{Lu2010}, performed EA for each subject individually, extracted the log-variance features, and finally used wAR \cite{drwuTHMS2017} for classification. This approach is denoted as TF-RCSP-EA-wAR in our analysis.

Again, the above TL pipeline did not re-reference the EEG channels. We also evaluated another slightly modified pipeline that performed CAR immediately after temporal filtering, which is denoted as TF-CAR-RCSP-EA-wAR.
\end{enumerate}

Figure~\ref{fig:TLresults} shows the TL performance of the six pipelines, averaged over 30 repeated runs to accommodate randomness. The last subfigure shows the overall TL performance, further averaged across the five subjects. We can observe that:

\begin{figure}[htpb] \centering
\includegraphics[width=\linewidth,clip]{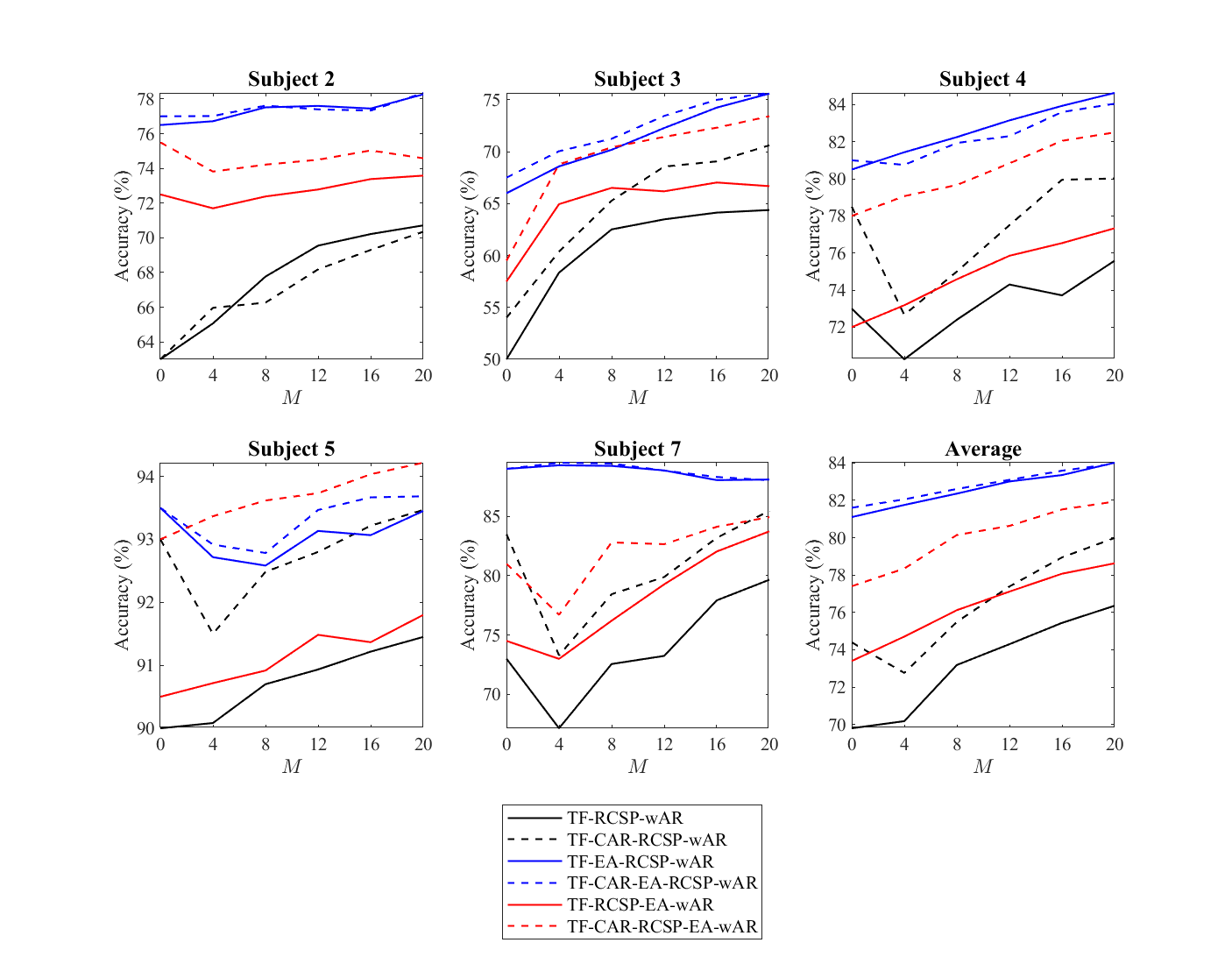}
\caption{The motor imagery classification performance of six different transfer learning pipelines. TF: Temporal filtering; RCSP: Regularized common spatial pattern; wAR: weighted adaptation regularization; CAR: Common average referencing; EA: Euclidean alignment.} \label{fig:TLresults}
\end{figure}

\begin{enumerate}
\item Performing EA, no matter whether it was between temporal filtering and spatial filtering, or after both temporal filtering and spatial filtering, always improved the TL performance, i.e., TF-EA-RCSP-wAR and TF-RCSP-EA-wAR always outperformed TF-RCSP-wAR, and TF-CAR-EA-RCSP-wAR and TF-CAR-RCSP-EA-wAR always outperformed TF-CAR-RCSP-wAR. This is because EA always reduces the marginal EEG data distribution discrepancies among the subjects, no matter where it is placed, and reducing these discrepancies always benefits TL.

\item Performing EA between temporal filtering and spatial filtering almost always achieved better performance than performing EA after both temporal filtering and spatial filtering, i.e., on average TF-EA-RCSP-wAR outperformed TF-RCSP-EA-wAR, and TF-CAR-EA-RCSP-wAR outperformed TF-CAR-RCSP-EA-wAR.

    EA should be placed after temporal filtering, because temporal filtering improves the EEG signal quality by smoothing it; particularly, it suppresses large EEG signal values, which may be outliers. This in turn improves the quality of the covariance matrices, and subsequently the EA reference matrix $\bar{R}$ and transformation matrix $\bar{R}^{-1/2}$.

    EA should be placed before spatial filtering, because RCSP utilizes EEG trials from both source and target subjects; EA reduces the marginal distribution discrepancies of EEG trials from different subjects, so that RCSP can identify spatial filters that work well for all subjects.

\item Performing CAR often led to performance improvement, i.e., TF-CAR-RCSP-wAR significantly outperformed TF-RCSP-wAR for four of the five subjects and on average, TF-CAR-EA-RCSP-wAR slightly outperformed TF-EA-RCSP-wAR on average, and TF-CAR-RCSP-EA-wAR always significantly outperformed TF-RCSP-EA-wAR. This is because CAR helps reduce the magnitude differences between EEG trials from different subjects.

\item For the best performing TF-CAR-EA-RCSP-wAR, the average binary classification accuracy between left hand and right hand for the five subjects, without any  labeled target-specific calibration data (i.e., $M=0$), was 81.60\%, and the classification accuracy increased very slowly with $M$ (e.g., the accuracy was 83.96\% when $M=20$). These suggest that EA exploits rich information from the source subjects, and enables plug-and-play motor imagery classification for a new subject.
\end{enumerate}
In summary, we confirmed through experiments that EA should be placed between temporal filtering and spatial filtering in traditional TL pipeline for motor imagery based BCIs, as in Figure~\ref{fig:TL}; and, performing CAR immediately after temporal filtering could further improve the performance.

\subsection{Where to Place EA in the BCI Deep TL Pipeline}

\begin{figure}[htpb] \centering
\includegraphics[width=.7\linewidth,clip]{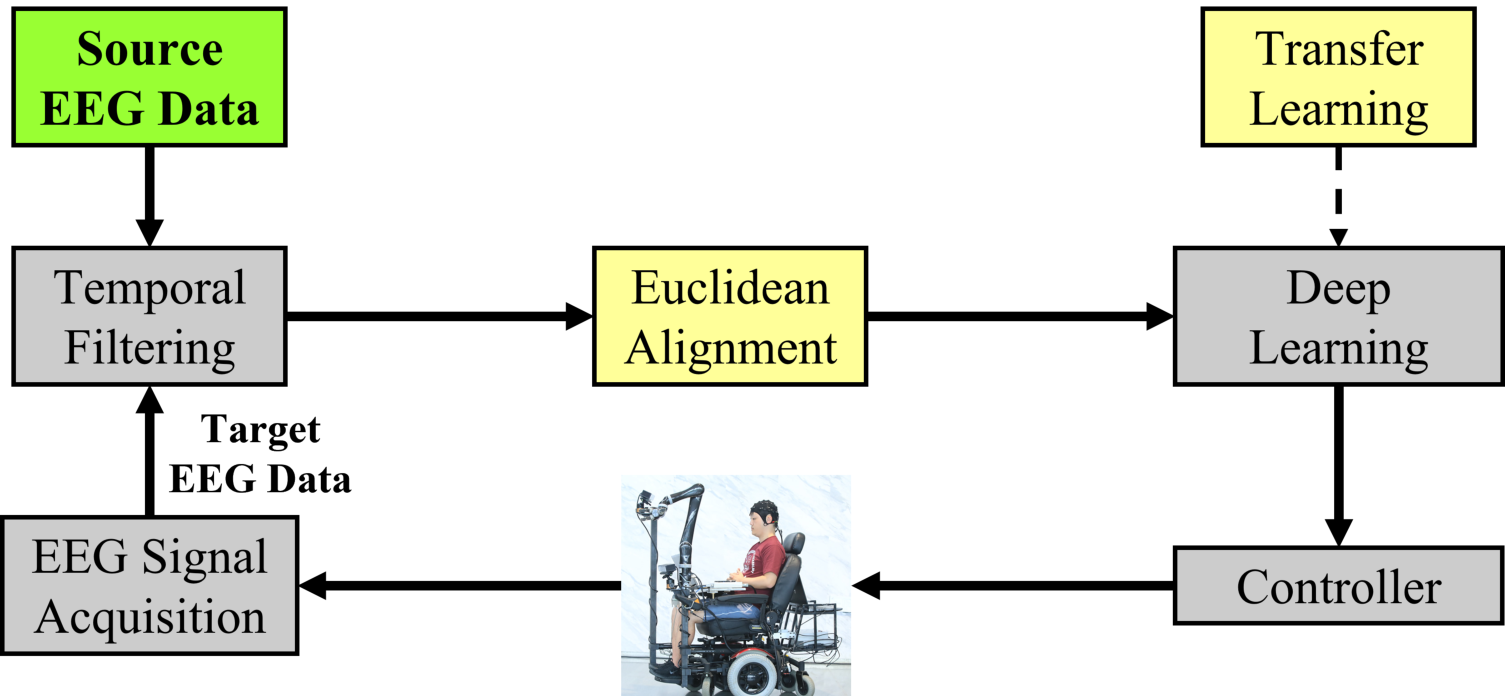}
\caption{The optimal location to perform Euclidean alignment in the BCI deep transfer learning pipeline (after temporal filtering).} \label{fig:DTL}
\end{figure}

When end-to-end deep learning approaches, e.g., EEGNet \cite{EEGNet}, are used, EA should also be placed between the temporal filtering block and the deep learning module, as illustrated in Figure~\ref{fig:DTL}. This subsection performs experiments to demonstrate that this placement indeed leads to better performance.

The dataset in the previous subsection was used again. We chose the popular EEGNet \cite{EEGNet} as the deep learning model, and compared three different configurations:
\begin{enumerate}
\item EEGNet-only, in which no band-pass temporal filtering and no EA were performed, i.e., the raw EEG signals were epoched into 3-second trials, and trials from all subjects were combined and fed directly into EEGNet.

\item EA-EEGNet, in which EA was performed without temporal filtering, i.e., the raw EEG signals from each subject were epoched into 3-second trials and aligned by EA, and then trials from all subjects were combined and fed directly into EEGNet.

\item TF-EA-EEGNet, as shown in Figure~\ref{fig:DTL}, in which the raw EEG signals from each subject were [8,30] Hz band-pass filtered, epoched into 3-second trials, aligned by EA, and then trials from all subjects were combined and fed into EEGNet.
\end{enumerate}

Due to much higher computational cost of EEGNet than traditional machine learning approaches, we performed only plug-and-play cross-subject validations, i.e., the target subject had no labeled calibration trials at all ($M=0$). The experiments for each subject were repeated three times to accommodate randomness, and the average results are reported in Table~\ref{tab:DTL}. Both EEGNet-only and EA-EEGNet had near-random binary classification accuracies, but TF-EA-EEGNet performed much better, suggesting the necessity to apply temporal filtering explicitly before EA in deep TL.

\begin{table*}[htbp]\centering \setlength{\tabcolsep}{4mm}
\caption{Classification accuracies (\%) of the three deep transfer learning configurations.} \label{tab:DTL}
\begin{tabular}{c|ccccc|c}
\toprule
Subject & 1 & 2 & 3& 4& 5& Average\\ \midrule
EEGNet-only & 49.83 & 50.00&48.00&49.00&49.83&49.33\\
EA-EEGNet   & 52.67 & 46.83&43.17&46.83&50.50&48.00\\
TF-EA-EEGNet& 70.17 & 59.83&59.50&91.50&66.33&66.33 \\ \bottomrule
\end{tabular}
\end{table*}

\subsection{Consistency of the EA Transformation Matrix Among Different Subjects}

The original EEG trials from different subjects have consistent channel configurations before alignment, because they use the same EEG headset. For example, for the MI trials in our experiments, the first channel was always AF3, the second channel was AF4, and the 59th channel was O2. It's important to maintain this channel consistency after EA, as channel location shifting may impact the subsequent classification performance.

\begin{figure}[htpb] \centering
\includegraphics[width=\linewidth,clip]{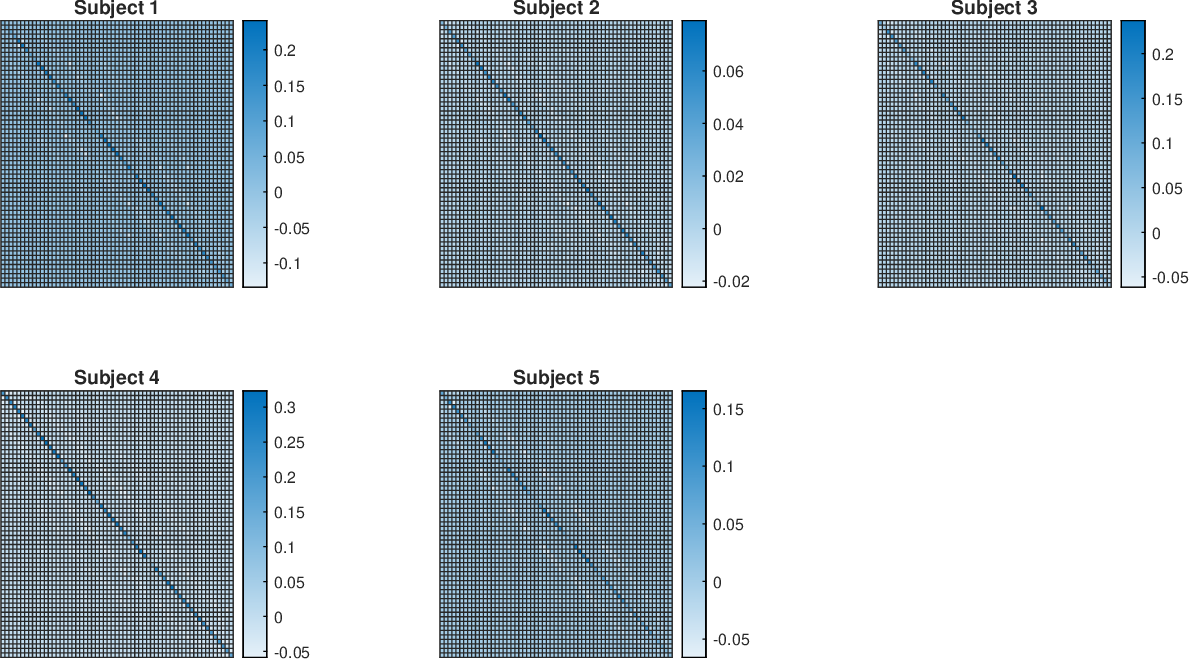}
\caption{Visualization of the EA transformation matrix $\bar{R}^{-1/2}$.} \label{fig:EAmatrix}
\end{figure}

Figure~\ref{fig:EAmatrix} uses heatmap to visualize the elements of the EA transformation matrix $\bar{R}^{-1/2}\in\mathbb{R}^{59\times 59}$ for the five subjects. There is a clear diagonal pattern, indicating that the original $i$-th channel in $X_n$ contributes the most to the transformed $i$-th channel in $\tilde{X}_n$. Hence, the channel consistency among different subjects is maintained after EA, although their transformation matrices are computed individually. This is another reason why EA achieves promising performance.

\subsection{Applications of EA}

In the original paper \cite{drwuEA2020}, the effectiveness of EA was only demonstrated in two classic BCI paradigms, motor imagery and P300 visual evoked potentials, using traditional classifiers. In the past five years, researchers from all over the world have applied it to 10 additional BCI paradigms, including mental imagery, rapid series visual presentation (RSVP), error-related negativity, sleep stage classification, motor+speech imagery, stress detection, emotion recognition, steady-state visual evoked potential (SSVEP), motor execution, and seizure detection, using both traditional classifiers and deep neural networks, all achieving promising performance improvements, as summarized in Table~\ref{tab:EA}. These 13 BCI paradigms cover almost all non-invasive BCI applications, i.e., EA has very broad applicability.

\begin{table*}[htbp]\centering \footnotesize \setlength{\tabcolsep}{2mm}
\caption{Classification accuracy improved by Euclidean alignment in 13 different BCI paradigms. Note that different publications generally used different datasets.} \label{tab:EA}
\begin{tabular}{c|c|c|c|c|c|c}
\toprule
\multirow{2.5}{*}{No.} &\multirow{2.5}{*}{BCI Paradigm} & \multirow{2.5}{*}{Classifier}& \multicolumn{3}{c|}{Accuracy (\%)} & \multirow{2.5}{*}{Reference}\\ \cmidrule(lr){4-6}
&& & w/o EA & w/ EA & Improvement & \\ \midrule
\multirow{39}{*}{1} &\multirow{39}{*}{Motor Imagery}
& CSP-LDA & 59.71 & 79.79 & \textbf{20.08} & He and Wu, IEEE TBME 2020 \cite{drwuEA2020} \\
&& CSP-LDA & 51.10  & 64.01  & \textbf{12.91} & He and Wu, IEEE TNSRE 2020 \cite{drwuLA2020} \\
&& TS-SVM & 56.65 & 64.32 & \textbf{7.67} & He and Wu, IEEE TNSRE 2020 \cite{drwuLA2020} \\
&& TIDNet & 61.53 & 70.44 & \textbf{8.91} & Kostas and Rudzicz, JNE 2020 \cite{Kostas2020} \\
&& CSP-LDA & 63.73 & 76.66 & \textbf{12.93} & Zhang and Wu, IEEE TNSRE 2020 \cite{drwuMEKT2020}\\
&& EEGInception & 63.72 & 72.44 & \textbf{8.72} & Bakas et al., NeurIPS 2021 \cite{Bakas2021NIPS} \\
&& CSP-LDA & 67.59 & 73.77 & \textbf{6.18} & Zhang et al., CMPB 2021 \cite{Zhang2021CMPB} \\
&& CSP-LDA & 67.75 & 73.53 & \textbf{5.78} & Liang et al., IROS 2021 \cite{Liang2021IROS} \\
&& CSP-SVM & 71.08 & 75.17 & \textbf{4.09} & Togha et al., BSPC 2021 \cite{Togha2021BSPC}  \\
&& AdaBN & 66.40 & 73.80 & \textbf{7.40} & Xu et al., JNE 2021 \cite{Xu2021JNE} \\
&& CSP-SVM & 65.78 & 75.56 & \textbf{9.78} & Demsy et al., IEEE SMC 2021 \cite{Demsy2021SMC} \\
&& TCA-W & 68.39 & 74.89 & \textbf{6.50} & Demsy, et al., IEEE SMC 2021 \cite{Demsy2021SMC} \\
&& CSP-LDA & 68.40 & 73.60 & \textbf{5.20} & Peterson et al., IEEE TBME 2021 \cite{Peterson2021TBME}  \\
&& EEGNet & 63.20 & 69.61 & \textbf{6.41} & Wu et al., NN 2022 \cite{drwuTLBCI2022} \\
&& ShallowCNN & 65.90 & 74.50 & \textbf{8.60} & Wu et al., NN 2022 \cite{drwuTLBCI2022} \\
&& CSP-LDA & 63.73 & 76.66 & \textbf{12.93} & Cai et al., JNM 2022 \cite{Cai2022JNM}  \\
&& CSP-LDA & 63.00 & 70.00 & \textbf{7.00} & Chen et al., JNUNS 2022 \cite{Chen2022JNUNS}  \\
&& NeuCube & 69.98 & 75.36 & \textbf{5.38} & Wu et al., NeuroComp. 2023 \cite{Wu2023NeuroCom} \\
&& EA-IISCSP & 79.75 & 81.93 & \textbf{2.18} & Wei and Ding, IEEE TNSRE 2023 \cite{Wei2023TNSRE} \\
&& RAVE & 79.20 & 82.90 & \textbf{3.70} & Pan et al., JNE 2023 \cite{Pan2023JNE} \\
&& CSP-LDA & 42.09 & 52.55 & \textbf{10.46} & Zhu et al., IEEE TIM 2023 \cite{Zhu2023TIM} \\
&& DSTL & 67.50 & 80.00 & \textbf{12.50} & Gao et al., IEEE TNSRE 2023 \cite{Gao2023TNSRE}  \\
&& MEIS & 59.11 & 66.91 & \textbf{7.80} & Liang et al., BSPC 2024 \cite{Liang2024BSPC}  \\
&& TLA-ACTL & 65.68 & 73.53 & \textbf{7.85} & Xu et al., BSPC 2024 \cite{Liang2024BSPC}  \\
&& TSL-SRT & 64.57 & 74.43 & \textbf{9.86} & Gao et al., CN 2024 \cite{Gao2024CogNeuro}   \\
&& CSP-LDA & 75.36 & 81.54 & \textbf{6.18} & Wang et al., KBS 2025 \cite{drwuTFT2025} \\
&& EEGNet & 43.80 & 48.90 & \textbf{5.10} & Bakas et al., JNE 2025 \cite{Bakas2025} \\
&& FBCSP & 52.62 & 79.98 & \textbf{27.36} & Amorim et al., BRACIS 2024 \cite{Amorim2024BRACIS} \\
&& EEGNet & 59.70 & 65.20 & \textbf{5.50} & Li et al., ArXiv 2025 \cite{Li2024spdim} \\
&& TSMNet & 60.00 & 65.30 & \textbf{5.30} & Li et al., ArXiv 2025 \cite{Li2024spdim} \\
&& TCNet & 74.38 & 77.70 & \textbf{3.32} & Wang et al., IEEE TBME 2025 \cite{Wang2024TBME}  \\
&& EEGNet & 69.70 & 72.10 & \textbf{2.40} & Wang et al., IEEE JBHI 2025 \cite{Wang2024JBHI} \\
&& CSP-LDA & 54.31 & 65.75 & \textbf{11.44} & Zhang et al., IEEE TIM 2024 \cite{Zhang2024TIM}  \\
&& FRCN & 57.64 & 70.41 & \textbf{12.77} & Xu et al., IEEE TIM 2024 \cite{Xu2024TIM}  \\
&& CSP-LDA & 57.55 & 67.06 & \textbf{9.51} & Zhu et al., RSI 2024 \cite{Zhu2024RSI} \\
&& CSP-LDA & 64.39 & 75.92 & \textbf{11.53} & Gao et al., NeuroComp. 2024 \cite{Gao2025NeuroCom}  \\
&& CSP-LDA & 67.45 & 71.44 & \textbf{3.99} & Jin et al., JNE 2024 \cite{Jin2024JNE}\\
&& SSCL-CSD & 64.15 & 67.32 & \textbf{3.17} & Li et al., JNE 2024 \cite{Li2024JNE} \\
&& BaseNet & 57.90 & 71.00 & \textbf{13.20} & Wimpff et al., arXiv \cite{Wimpff2025} \\ \midrule
2&Mental Imagery & MDM & 56.02 & 66.08 & \textbf{10.06} & Kumar et al., BCI 2019 \cite{Kumar2019BCI}  \\ \midrule
3&Event-Related Potential & xDAWN-SVM & 64.60 & 68.80 & \textbf{4.20} & He and Wu, IEEE TBME 2020 \cite{drwuEA2020} \\ \midrule
4&RSVP & CSP-LDA & 65.36 & 69.07 & \textbf{3.71} & \multirow{2}{*}{Zhang and Wu, IEEE TNSRE 2020 \cite{drwuMEKT2020}} \\ \cline{1-6}
5&Error-Related Negativity & CSP-LDA & 61.87 & 64.63 & \textbf{2.76} & \\ \midrule
6&Sleep Stage Classification & EEGInception & 56.87 & 67.70 & \textbf{10.83} & Bakas et al., NeurIPS 2021 \cite{Bakas2021NIPS}  \\ \midrule
\multirow{2}{*}{7}&\multirow{2}{*}{Motor+Speech Imagery} & DS-CSP & 63.62 & 67.09 & \textbf{3.47} & \multirow{2}{*}{Zhan et al., CBM 2022 \cite{Zhan2022}} \\
&& DS-TSM & 66.12 & 68.42 & \textbf{2.30} &  \\ \midrule
8&Stress Detection & SDCAN & 45.69 & 48.17 & \textbf{2.48} & Fu et al., IEEE TNSRE 2022 \cite{Fu2022TNSRE}  \\ \midrule
9&Emotion Recognition & AM-MSFFN & 88.00 & 93.51 & \textbf{5.51} & Jiang et al., IEEE SJ 2023 \cite{Jiang2023IEEESJ} \\ \midrule
10&SSVEP & SUTL & 72.28 & 74.04 & \textbf{1.76} & Li et al., ESWA 2024 \cite{Li2024ESWA} \\ \midrule
11&Motor Execution & EEGNet & 52.10 & 62.50 & \textbf{10.40} & Bakas et al., JNE 2025 \cite{Bakas2025} \\ \midrule
12&Seizure Detection & EEGNet & 68.43 & 75.86 & \textbf{7.43} & Wang et al., NSR 2025 \cite{drwuNSR2025} \\ \midrule
13&\tabincell{c}{Traumatic Brain\\ Injury Classification} & CNN & 61.00 & 89.00 & 28.00 & Vishwanath et al., BSPC 2025 \cite{Vishwanath2025} \\ \bottomrule
\end{tabular}
\end{table*}
\normalsize

EA was also given very high evaluations in the literature.

For example, Kostas and Rudzicz \cite{Kostas2020} from the University of Toronto integrated EA, mixup data augmentation with their proposed TIDNet for EEG-based motor imagery classification, and found that ``\emph{while TIDNet alone or with only mixup shows mixed results, once alignment is included there is a nearly universal performance benefit to the deeper network (TIDNet), this improvement is much more consistent when both mixup and EA are used, with only a single (statistically insignificant) case showing very mildly worse performance. ... Focusing on the largest increase in performance, we find a significant 8.91\% increase in 4-way classification performance using EA, jumping from 61.53\% to 70.44\%. ... in terms of both trend and peak performance, EA seems consistently beneficial to EEGNet as well as TIDNet. ... Our empirical results are clear that using both (EA and mixup) consistently increases performance but, in terms of magnitude of increase, EA alone is many times better performing.}"

Junqueira et al. \cite{Junqueira2024} performed comprehensive experiments and comparisons on using EA in deep learning based motor imagery classification in their Journal of Neural Engineering publication entitled ``A systematic evaluation of Euclidean alignment with deep learning for EEG decoding." They found that ``\emph{EA improved the mean accuracy for all cross-subject models and datasets evaluated and led to a 70\% acceleration in convergence in the shared models. Consequently, we believe that EA should be a standard pre-processing step when training cross-subject models.}"

EA has also achieved remarkable performance in BCI competitions.

By integrating it into our TL algorithms, the author's team won nine National Championships in China BCI Competition (the most prestigious BCI competition in China, organized jointly by the National Natural Science Foundation of China, Chinese Institute of Electronics, and Tsinghua University) between 2019 and 2024.

In the Benchmarks for EEG Transfer Learning\footnote{https://beetl.ai/challenge} (BEETL) competition organized at NeurIPS 2021, the champion team's approach \cite{Wei2022} ``\emph{for improving subject-independence is to align the latent feature distributions of the deep learning models between different subjects and sessions. This is similar to the Euclidean Alignment method, which performs spatial whitening with the average covariance matrix per subject, although the methods are not constrained to be applied on the model input.}" In the runner-up's approach \cite{Wei2022}, ``\emph{Euclidean Alignment is used to close the gap between subjects in source domains and target domain. All trials of each subject are aligned such that the mean covariance matrices of all subjects are equal to the identity matrix after alignment. After alignment, data set distribution in both source domains and target domain are more similar.}"

\section{Extension/Counterparts of EA} \label{sect:ext}

This section introduces LA, an extension of the unsupervised EA to supervised EEG data alignment, and several other EEG data alignment approaches.

\subsection{Label Alignment}

LA \cite{drwuLA2020} was also proposed by the author's group in 2020, primarily for heterogeneous EEG-based motor imagery classification, i.e., the motor imagery classes for different subjects are partially or completely different. LA assumes all source domain samples are labeled, and the target domain also has a small number of labeled calibration samples. The label information could be properly used for better EEG trial alignment.

The idea of LA is shown in Figure~\ref{fig:EALA}. It essentially performs class-wise EA to reduce the class conditional  EEG data distribution discrepancies between the source and target domains.

As an example, assume the source subject imagines the movement of left hand or right hand, whereas the target subject imagines the movement of left hand or both feet. Then, we pair the source subject left hand motor imagery movement with the target subject left hand movement, and the source subject right hand movement with the target subject both feet movement. For a given class-pair, assume the source domain has $N_\mathrm{s}$ labeled EEG trials $\{(X_n^\mathrm{s},y_n^\mathrm{s})\}_{n=1}^{N_\mathrm{s}}$, and the target domain has $N_\mathrm{t}$ labeled EEG trials $\{(X_n^\mathrm{t},y_n^\mathrm{t})\}_{n=1}^{N_\mathrm{t}}$.

LA first computes class-wise mean covariance matrices $\bar{R}_\mathrm{s}$ and $\bar{R}_\mathrm{t}$:
\begin{align}
  \bar{R}_\mathrm{s}&=\frac{1}{N_\mathrm{s}}\sum_{n=1}^{N_\mathrm{s}} X_n^\mathrm{s} (X_n^\mathrm{s})^\top, \\
  \bar{R}_\mathrm{t}&=\frac{1}{N_\mathrm{t}}\sum_{n=1}^{N_\mathrm{t}} X_n^\mathrm{t} (X_n^\mathrm{t})^\top.
\end{align}
Note that both $\bar{R}_\mathrm{s}$ and $\bar{R}_\mathrm{t}$ are symmetric and semi-positive definite.

LA then performs the alignment only to the source domain EEG trials:
\begin{align}
  \tilde{X}_n^\mathrm{s}=\bar{R}_\mathrm{t}^{1/2}\bar{R}_\mathrm{s}^{-1/2}X_n^\mathrm{s},\quad n=1,...,N_\mathrm{s}.
\end{align}
Here $\bar{R}_\mathrm{t}^{1/2}\bar{R}_\mathrm{s}^{-1/2}$ is called the transformation matrix in LA.

After LA, the class-wise mean covariance matrix of all $N_\mathrm{s}$ aligned source domain trials $\{\tilde{X}_n^\mathrm{s}\}_{n=1}^{N_\mathrm{s}}$ equals the paired class-wise mean covariance matrix of the $N_\mathrm{t}$ target domain trials $\{\tilde{X}_n^\mathrm{t}\}_{n=1}^{N_\mathrm{t}}$:
\begin{align}
\frac{1}{N_\mathrm{s}}\sum_{n=1}^{N_\mathrm{s}}\tilde{X}_n^\mathrm{s}(\tilde{X}_n^\mathrm{s})^\top  &=\frac{1}{N_\mathrm{s}}\sum_{n=1}^{N_\mathrm{s}}\bar{R}_\mathrm{t}^{1/2} \bar{R}_\mathrm{s}^{-1/2} X_n^\mathrm{s}
(X_n^\mathrm{s})^\top\bar{R}_\mathrm{s}^{-1/2}\bar{R}_\mathrm{t}^{1/2}\nonumber \\
&=\bar{R}_\mathrm{t}^{1/2}\bar{R}_\mathrm{s}^{-1/2} \bar{R}_\mathrm{s} \bar{R}_\mathrm{s}^{-1/2}\bar{R}_\mathrm{t}^{1/2}\nonumber\\
&=\bar{R}_\mathrm{t}^{1/2} \bar{R}_\mathrm{t}^{1/2}\nonumber\\
&=\bar{R}_\mathrm{t}\nonumber\\
&=\frac{1}{N_\mathrm{t}}\sum_{n=1}^{N_\mathrm{t}}X_n^\mathrm{t}(X_n^\mathrm{t})^\top
\end{align}
i.e., for each paired class, the mean covariance matrices of both domain are equal, and hence the class conditional EEG data distributions from different domains become more consistent.

LA performs the above alignment for each paired class in the source domain separately. Note that it transforms the source domain trials to match the target domain trials. Unlike in EA that all EEG trials in the same domain share the same transformation matrix, in LA different classes in the same domain use different transformation matrices. When LA is used in testing, we do not know the class label (this is to be predicted) for an incoming target domain EEG trial, so we cannot determine which transformation matrix to use. By transforming the known source domain trials only, we can avoid this problem.

We \cite{drwuLA2020} showed that LA can also be used when the source and target domains have completely different label spaces. For example, when the source domain has two classes of left hand and right hand, and the target domain has two classes of feet and tongue, we can align the source domain left hand class with the target domain feet class, and the source domain right hand class with the target domain tongue class, still achieving effective TL. Of course, EA could also be used in this challenging scenario, as it does not distinguish among the classes at all.

LA has demonstrated promising performance in multiple BCI paradigms, particularly in heterogeneous TL, including motor imagery classification \cite{drwuLA2020,Wei2022,Xu2023,Ma2024}, motor+speech imagery classification \cite{Zhan2022}, traumatic brain injury classification \cite{Vishwanath2022}, emotion recognition \cite{Peng2023,Ren2024,Zhang2024}, and mental workload classification \cite{Wang2024}.

\begin{table*}[htbp]\centering  \setlength{\tabcolsep}{2mm}
\caption{Classification accuracy improved by label alignment in 5 different BCI paradigms. Note that different publications generally used different datasets.} \label{tab:LA}
\begin{tabular}{c|c|c|c|c|c|c}
\toprule
\multirow{2.5}{*}{No.} &\multirow{2.5}{*}{BCI Paradigm} & \multirow{2.5}{*}{Classifier}& \multicolumn{3}{c|}{Accuracy (\%)} & \multirow{2.5}{*}{Reference}\\ \cmidrule(lr){4-6}
&& & w/o LA & w/ LA & Improvement & \\ \midrule
\multirow{2}{*}{1}&\multirow{2}{*}{\tabincell{c}{Motor \\ Imagery}} & CSP-LDA & 51.10 & 69.95 & \textbf{18.85} & \multirow{2}{*}{He and Wu, IEEE TNSRE 2020 \cite{drwuLA2020}} \\
&& TS-SVM & 56.65 & 71.75 & \textbf{15.10} &  \\ \midrule
\multirow{2}{*}{2}&\multirow{2}{*}{\tabincell{c}{Motor+Speech\\ Imagery}} & DS-CSP & 63.62 & 66.64 & \textbf{3.02} & \multirow{2}{*}{Zhan et al., CBM 2022 \cite{Zhan2022}} \\
&& DS-TSM & 66.12 & 71.12 & \textbf{5.00} &  \\ \midrule
\multirow{6}{*}{3}&\multirow{6}{*}{\tabincell{c}{Traumatic\\ Brain Injury\\ Classification}} & DT & 62.24 & 65.78 & \textbf{3.54} & \multirow{6}{*}{Vishwanath et al., IEEE EMBC 2022 \cite{Vishwanath2022}} \\
&& RF & 67.93 & 70.18 & \textbf{2.25} &  \\
&& kNN & 71.31 & 74.42 & \textbf{3.11} &  \\
&& MLP & 73.66 & 78.73 & \textbf{5.07} &  \\
&& SVM & 74.46 & 74.86 & \textbf{0.40} &  \\
&& XGB & 66.89 & 69.41 & \textbf{2.52} &  \\ \midrule
\multirow{3}{*}{4}&\multirow{3}{*}{\tabincell{c}{Emotion\\ Recognition}} & TCA& 46.72 & 68.08 & \textbf{21.36} & Peng et al., IEEE TII 2023 \cite{Peng2023}  \\
&& kNN  & 48.14 & 66.46 & \textbf{18.32} & Ren et al., KBS 2024 \cite{Ren2024}  \\
&& GFK  & 42.80 & 68.10 & \textbf{25.30} & Zhang et al., IEEE JBHI 2024 \cite{Zhang2024}  \\ \midrule
5&\tabincell{c}{Mental\\ Workload\\ Classification} & LDA & 56.63 & 82.53 & \textbf{25.90} & Wang et al., IEEE JBHI 2024 \cite{Wang2024} \\ \bottomrule
\end{tabular}
\end{table*}

LA may outperform EA when the source and target domains have different label spaces. For example, we \cite{drwuLA2020} demonstrated that when the source subject imagines the movements of right hands or both feet, whereas the target subject imagines the movements of left hand or tongue, the baseline CSP-LDA (linear discriminant analysis) classification accuracy without any alignment was 51.10\%, the classification accuracy of EA-CSP-LDA was 64.01\%, and the classification accuracy of LA-CSP-LDA was 69.95\%.

\begin{figure}[htpb] \centering
\includegraphics[width=\linewidth,clip]{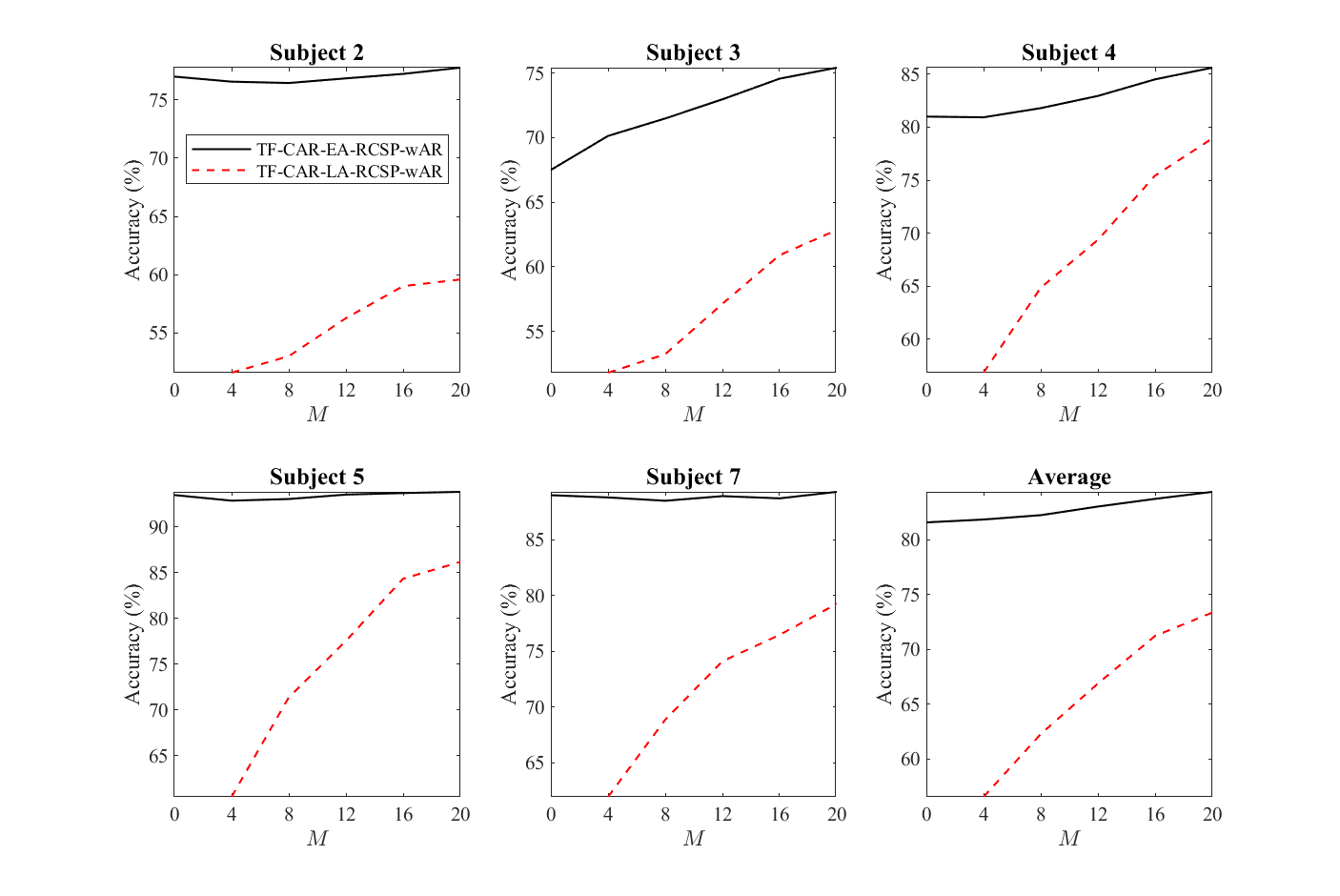}
\caption{The motor imagery classification performance of six different transfer learning pipelines. TF: Temporal filtering; RCSP: Regularized common spatial pattern; wAR: weighted adaptation regularization; CAR: Common average referencing; EA: Euclidean alignment; LA: Label alignment.} \label{fig:EALAresults}
\end{figure}

However, when the source and target domains have the same label space, EA may outperform LA. For example, Figure~\ref{fig:EALAresults} shows the performance comparison of TF-CAR-EA-RCSP-wAR (the best-performer in Figure~\ref{fig:TLresults}) and TF-CAR-LA-RCSP-wAR, which replaces EA by LA. LA performed much worse than EA, because given a small number of labeled target domain samples, it is difficult to accurately estimate $\bar{R}_\mathrm{t}^{1/2}$ in the target domain, and hence the transformation matrix $\bar{R}_\mathrm{t}^{1/2}\bar{R}_\mathrm{s}^{-1/2}$. He and Wu \cite{drwuLA2020} used $k$-medoids clustering to select $k$ most representative unlabeled EEG trials to label, making the estimation of $\bar{R}_\mathrm{t}^{1/2}$ more reliable than random selection, improving the performance of LA.

In summary, when the source and target domains have the same label space, EA is preferred over LA; otherwise, LA is preferred over EA.

\subsection{Comparison of EEG Data Alignment Approaches}

In addition to Riemannian alignment, EA and LA, there are several other EEG data alignment approaches proposed in the literature,  which can be categorized into two classes:
\begin{enumerate}
\item Alignments of the EEG trials by transformation matrices, and the aligned EEG trials can then be used in both traditional machine learning and deep learning. Representative approaches include Riemannian  Alignment \cite{Zanini2018}, Parallel Transport \cite{Yair2019}, Riemannian Procrustes Analysis \cite{Rodrigues2019}, EA \cite{drwuEA2020}, Centroid Alignment \cite{drwuMEKT2020}, and LA \cite{drwuLA2020}.

\item Alignments of the mini-batches in deep learning, which are specific to deep learning. Representative approaches include Adaptive Batch Normalization (adaBN) \cite{JimenezGuarneros2020,Xu2021} and Latent Alignment \cite{Bakas2025}.
\end{enumerate}

Table~I in our previous publication \cite{drwuTLBCI2022} gives a comprehensive comparison of the first category of EEG data alignment approaches. For the completeness of this paper, we collect and update its results, and add adaBN and latent alignment, as shown in Table~\ref{tab:alignment}.

\begin{table*}[htbp]  \centering \footnotesize \setlength{\tabcolsep}{.5mm}
\caption{Comparison of different EEG data alignment approaches \cite{drwuTLBCI2022}.}   \label{tab:alignment}
\begin{tabular}{c||c|c|c|c|c|c|c|c}
\toprule
Criterion & \tabincell{c}{Riemannian\\ Alignment\\ \cite{Zanini2018}} & \tabincell{c}{Parallel\\ Transport \\ \cite{Yair2019}} & \tabincell{c}{Riemannian\\ Procrustes\\ Analysis \cite{Rodrigues2019}} & \tabincell{c}{Euclidean \\ Alignment\\ \cite{drwuEA2020}}  & \tabincell{c}{Centroid\\ Alignment\\ \cite{drwuMEKT2020}} & \tabincell{c}{Label\\ Alignment\\ \cite{drwuLA2020}} & \tabincell{c}{adaBN \\ \cite{JimenezGuarneros2020,Xu2021}}& \tabincell{c}{Latent \\ Alignment\\ \cite{Bakas2025}}\\ \midrule

\tabincell{c}{Applicable\\ Paradigms}	& \tabincell{c}{7 different\\ paradigms\\ (Table~\ref{tab:RA})} &	MI \cite{Yair2019} &	\tabincell{c}{MI \cite{Rodrigues2019},\\ ERP \cite{Rodrigues2019},
\\ SSVEP \cite{Rodrigues2019}} &\tabincell{c}{13 different \\ paradigms \\ (Table~\ref{tab:EA})}	& \tabincell{c}{MI \cite{drwuMEKT2020},\\ ERP \cite{drwuMEKT2020}, \\ Emotion\\  recognition\\ \cite{Peng2023,Ren2024,Zhang2024}} &	\tabincell{c}{5 different \\ paradigms \\ (Table~\ref{tab:LA})} & \tabincell{c}{MI \cite{Xu2021}, \\ Cognitive load\\
recognition\\ \cite{JimenezGuarneros2020}}& \tabincell{c}{MI \cite{Bakas2025}, \\ ERP \cite{Bakas2025}, \\ Sleep stage\\ classifica-\\tion \cite{Bakas2025}}\\\midrule

\tabincell{c}{Online or\\ Offline} &	Both	& Both &	Both & Both &	Both &	Offline & Both & Both\\\midrule

\multicolumn{1}{c||}{\tabincell{c}{Need\\ Labeled Target\\ Domain Trials }} &	\multicolumn{1}{|c|}{\tabincell{c}{No for MI,\\ Yes for others}}	& No &	Yes	& No &	No &	Yes& No & No\\\midrule

\tabincell{c}{What to\\ Align} &	
\multicolumn{1}{|c|}{\tabincell{c}{Riemannian\\ space \\ covariance\\ matrices}} &	\multicolumn{1}{|c|}{\tabincell{c}{Riemannian\\ Tangent \\space\\ features}}	& \multicolumn{1}{|c|}{\tabincell{c}{Riemannian\\ space\\ covariance\\ matrices}}	& \multicolumn{1}{|c|}{\tabincell{c}{Euclidean\\ space\\ EEG\\ trials}} &	\multicolumn{1}{|c|}{\tabincell{c}{Riemannian\\ space\\ covariance\\ matrices}} &	\multicolumn{1}{|c|}{\tabincell{c}{Euclidean\\ space\\ EEG\\ trials}}& \multicolumn{1}{|c|}{\tabincell{c}{Euclidean\\ space EEG\\ features\\ in a\\ mini-batch}} & \multicolumn{1}{|c}{\tabincell{c}{Euclidean\\ space EEG\\ features\\ in a\\ mini-batch}}\\\midrule

\multicolumn{1}{c||}{\tabincell{c}{Reference\\ Matrix \\ Calculation}} &	
\multicolumn{1}{|c|}{\tabincell{c}{Riemannian\\ mean of \\ resting or \\ non-target\\ trial\\  covariance\\ matrices in\\ each domain}}	&
\multicolumn{1}{|c|}{\tabincell{c}{Riemannian\\ mean of all\\ covariance \\matrices in \\each domain}} &	\multicolumn{1}{|c|}{\tabincell{c}{Riemannian\\ mean of all \\ labeled\\ covariance \\matrices in \\each domain}}	&
\multicolumn{1}{|c|}{\tabincell{c}{Euclidean\\ mean of all\\ covariance \\matrices in \\each domain}}	& \multicolumn{1}{|c|}{\tabincell{c}{Riemannian,\\ Euclidean, or\\ Log-Euclidean\\ mean of all\\ covariance\\ matrices in\\ each domain}} &	
\multicolumn{1}{|c|}{\tabincell{c}{Log-Euclidean\\ mean of\\ labeled \\covariance\\ matrices \\ in each class \\of each domain}} & Unneeded & Unneeded\\\midrule

Classifier	& \multicolumn{1}{|c|}{\tabincell{c}{Riemannian\\ space only}}	& \multicolumn{1}{|c|}{\tabincell{c}{Euclidean\\ space only}}	& \multicolumn{1}{|c|}{\tabincell{c}{Riemannian \\space only}}	& \multicolumn{1}{|c|}{\tabincell{c}{Riemannian or\\ Euclidean space}}	& \multicolumn{1}{|c|}{\tabincell{c}{Riemannian or\\ Euclidean space}}	& \multicolumn{1}{|c|}{\tabincell{c}{Riemannian or\\ Euclidean space}}&
\multicolumn{1}{|c}{\tabincell{c}{Euclidean\\ space only}} & \multicolumn{1}{|c}{\tabincell{c}{Euclidean\\ space only}}\\\midrule

\multicolumn{1}{c||}{\tabincell{c}{Handle Class\\ Mismatch\\ between\\ Domains}} &	No &	No&	No&	No&	No&	Yes & No & No\\\midrule

\tabincell{c}{Computational\\ Cost}	&High	&High&	High&	Low&	Low&	Low & Low & Low\\ \bottomrule
\end{tabular}
\end{table*}

In application domains other than BCIs, many classic data alignment approaches for TL have also been proposed \cite{Zhuang2021}, e.g., transfer component analysis (TCA) \cite{Pan2011b} and CORrelation ALignment (CORAL) \cite{Sun2016}.

TCA assumes that both marginal and conditional probability distributions of the source domain $X_\mathrm{s}$ and target domain $X_\mathrm{t}$ are different, and tries to find a transformation matrix $W$ to make them approximately identical. More specifically, $W$ needs to satisfy two constraints \cite{Pan2011b}: 1) the distance between the marginal distributions of $WX_\mathrm{s}$ and $WX_\mathrm{t}$ is small; and, 2) $WX_\mathrm{s}$ and $WX_\mathrm{t}$ preserve important properties of $X_\mathrm{s}$ and $X_\mathrm{t}$. Compared with TCA, EA only considers the first constraint but not the second, which could be a direction for further improvement.

CORAL minimizes the domain shift by aligning the second-order statistics of their distributions. Assume a $d$-dimensional feature vector has been extracted for each sample in each domain. CORAL first computes the feature covariance matrices $C_\mathrm{s}\in\mathbb{R}^{d\times d}$ and $C_\mathrm{t}\in\mathbb{R}^{d\times d}$ in the source domain and target domain, and then finds a transformation matrix $W\in\mathbb{R}^{d\times d}$ for the source domain such that the Frobenius norm of the difference between the covariance matrices, $||W^\top C_\mathrm{s}W-C_\mathrm{t}||_F^2$, is minimized. As pointed out in \cite{drwuEA2020,drwuLA2020}, the main difference between CORAL and EA/LA is that: ``\emph{CORAL uses 1D features, and each domain has only one feature covariance matrix, which measures the covariances between different pairs of individual features. EA/LA uses 2D features (EEG trials), and each EEG trial has a covariance matrix, which measures the covariances between different pairs of EEG channels. So, the covariance matrices in CORAL and EA/LA have different meanings.}" Meanwhile, CORAL may also be used to align the log-variance features in TF-CAR-EA-RCSP-wAR before they are fed into wAR for classification.

\section{New Research Directions} \label{sect:future}

This section introduces some potential new research directions based on EA.

\subsection{Applicability of EA to Other Multi-Channel Physiological Signals}

The original EA was proposed for aligning multi-channel EEG signals from different subjects/sessions. It may also be used in aligning other multi-channel physiological signals.

For example, Wang et al. \cite{drwuNSR2025} used EA as a standard preprocessing step on two intracranial EEG datasets (eight humans and four canines in the Kaggle UPenn and Mayo Clinic's Seizure Detection Challenge \cite{Baldassano2017}, and 21 human patients in the Freiburg dataset \cite{Ihle2012}), to address the heterogeneities in electrode placements across subjects, and hence to enable cross-species and cross-modality epileptic seizure detection.

Since many other physiological signals, e.g., Magnetoencephalography (MEG), Electrocardiography (ECG) and stereo-Electroencephalography (sEEG), also have multiple channels and large individual differences, it is interesting to study if EA can also be applied to them to reduce the data distribution discrepancies among subjects.

\subsection{Large Models for EEG Decoding}

As large models have achieved great successes in natural language processing and computer vision, they have also started to find applications in EEG-based BCIs \cite{Jiang2024,Zhang2024a}. For example, Jiang et al. \cite{Jiang2024} showed that their large brain model, pre-trained on about 2,500 hours of various types of EEG signals from around 20 datasets, outperformed all compared SOTA in EEG-based abnormal detection, event type classification, emotion recognition, and gait prediction.

However, Jiang et al. \cite{Jiang2024} also pointed out that one challenge in training large models for EEG decoding is the lack of sufficient EEG data. This could be alleviated by making use of relevant data from other species and/or modalities, as our recent work \cite{drwuNSR2025} has shown that canines's scalp/intracranial EEG data could be used to facilitate scalp/intracranial EEG based seizure classification for humans, and vice versa. However, as mentioned in the previous subsection, this was impossible without using EA to align the trials.

Another important consideration in training large models for EEG-based BCIs is the efficiency. Without speaking, the faster the better. Junqueira et al. \cite{Junqueira2024} pointed out in their evaluation of EA with deep learning for EEG decoding that ``\emph{when using single shared models with training data from all individuals, using EA allowed us to achieve the same level of accuracy as the models without EA using 70\% fewer iterations and a final level of accuracy 4.33\% higher when using the same number of iterations.}" Since large models are also deep neural network based, EA could be used as a pre-processing step to align EEG data from different subjects/sessions/species/modalities to accelerate the training.

\subsection{Extension from Classification to Regression}

So far, the applications of EA exclusively focused on BCI classification problems, such as motor imagery, event related potential, seizure classification, etc., as shown in Table~\ref{tab:EA}. However, there are also many important regression problems in BCIs \cite{drwuTLBCI2022}, such as primitive emotion estimation \cite{drwuPIEEE2023}, driver drowsiness estimation \cite{drwuFWET2019}, user reaction time estimation \cite{drwuRG2017}, etc. As in BCI classification problems, EEG signal stationarity and large individual differences also exist in BCI regression problems. So, TL could also be used to facilitate the calibration \cite{drwuTFS2017}.

Our previous research has extended some blocks in the TL pipeline for motor imagery classification (Figure~\ref{fig:TL}), e.g., common spatial pattern filtering \cite{drwuSF2018} and wAR classifier \cite{drwuTFS2017}, to BCI regression problems, using fuzzy sets. It is interesting to study if EA/LA could be extended from classification to regression, also using fuzzy sets.

\subsection{Accurate and Robust EEG Decoding}

Most BCI decoding research so far only focused on accurate decoding of the brain signals. However, studies have shown that the decoding algorithms may suffer from adversarial attacks \cite{drwuBCIAttack2019,drwuTAR2019,drwuNSO2022,drwuSCIS2022,drwuAP2023,drwuBackdoor2023,drwuIF2024}, i.e., a small perturbation, which may be too small to be detected by human eyes or computer algorithms, can be added to the benign EEG signal to mislead the decoding algorithm.

Various adversarial attack approaches have been proposed for motor imagery classification \cite{drwuBCIAttack2019,drwuAP2023,drwuBackdoor2023,drwuIF2024}, P300 event related potential classification \cite{drwuBCIAttack2019,drwuAP2023,drwuBackdoor2023,drwuIF2024}, feedback error-related negativity classification \cite{drwuAP2023,drwuBackdoor2023,drwuIF2024}, steady-state visual evoked potential classification \cite{drwuSCIS2022,drwuNSR2021}, and also regression problems \cite{drwuTAR2019}. They could cause various forms of damage. For example, attacking a BCI speller could mis-interpret the user's opinion, and attacking a motor imagery BCI controlled wheelchair could intentionally drive the user into danger. More seriously, as pointed out in \cite{RAND2020}, ``\emph{adversary hacking into BCI devices that influence the motor cortex of human operators could theoretically send false directions or elicit unintended actions, such as friendly fire.}"

Thus, it is urgent to consider how to defend against adversarial attacks to BCIs. As demonstrated in our benchmark studies \cite{drwuFGCS2023}, a promising approach is adversarial training, i.e., to add adversarial samples into the training dataset so that the decoding algorithm is less vulnerable to them in testing. However, while increasing the robustness on adversarial samples, adversarial training usually deteriorates the decoding performance on the benign samples, as the training dataset is polluted. It is desirable to increase the decoding performance on both benign and adversarial EEG trials, i.e., to achieve simultaneously accurate and robust decoding.

\begin{figure}[htpb] \centering
\includegraphics[width=\linewidth,clip]{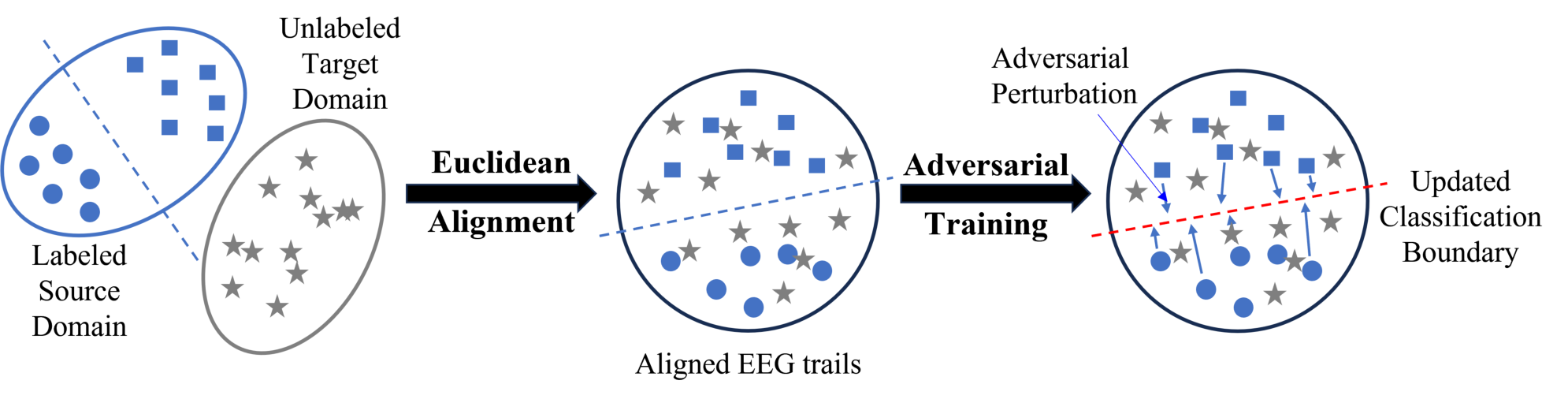}
\caption{Alignment based adversarial training (ABAT) to simultaneously increase the decoding accuracy and adversarial robustness.} \label{fig:ABAT}
\end{figure}

Our recently proposed alignment based adversarial training (ABAT) algorithm \cite{drwuABAT2024} solved this challenging problem. Its flowchart is shown in Figure~\ref{fig:ABAT}. Essentially, it first uses EA to reduce the data distribution discrepancies among different domains, and hence to improve the classification accuracy on benign EEG trials; it then performs adversarial training to improve the robustness on adversarial samples.

More recently, we \cite{drwuBenchmark2025} showed that simultaneously integrating EA, data augmentation and adversarial training can further improve both the accuracy and robustness of EEG-based BCIs.

In summary, EA is essential in adversarial training to simultaneously improve the decoding accuracy and adversarial robustness of EEG-based BCIs. It is expected that more sophisticated and comprehensive TL pipeline, like the one shown in Figure~\ref{fig:TL}, can be integrated with adversarial training to achieve better performance.

\subsection{Accurate, Robust, and Privacy-Preserving EEG Decoding}

In addition to adversarial security, privacy protection is another important ethic consideration in BCIs. Our survey \cite{drwuTCSS2023} shows that many types of private information, e.g., personal identify, age, emotions, brain-related diseases, etc., could be mined from EEG signals. Multiple laws and regularization all over the world, particularly the General Data Protection Regularization (GDPR) \cite{GDPR2016} of the European Union, and the Personal Information Protection Law of China, impose strict privacy protection requirements. Ideally, for privacy-preserving TL, there should be no sharing of authentic EEG data among the source domains, and also between the source domains and the target domain.

Various approaches, including security multiparty computation \cite{drwuPrivacy2019}, synthetic data generation \cite{Debie2020,Pascual2021}, homomorphic encryption \cite{Popescu2021}, source-free TL \cite{drwuTBME2022,drwuMSDT2022,drwuSFDA2023,drwuLSFT2023}, federated learning \cite{drwuFed2024}, machine unlearning \cite{Shao2024}, and data perturbation \cite{drwuPrivacySMC2024,drwuPertub2025}, have been proposed for privacy-preserving BCIs.

A very challenging problem is how to simultaneously achieve accurate decoding, adversarial robustness and privacy-protection in EEG-based BCIs. To our knowledge, aligned and augmented adversarial ensemble (A3E) \cite{drwuA3E2024} was the only solution so far, where EA also plays a very important role.

The flowchart of using A3E in EEG-based motor imagery classification is shown in Figure~\ref{fig:A3E}. Each source domain uses EA to align their EEG trials first. Since EA uses information from its own domain only, there is no privacy issues. Three privacy protection scenarios are then considered:

\begin{figure}[htpb] \centering
\includegraphics[width=.8\linewidth,clip]{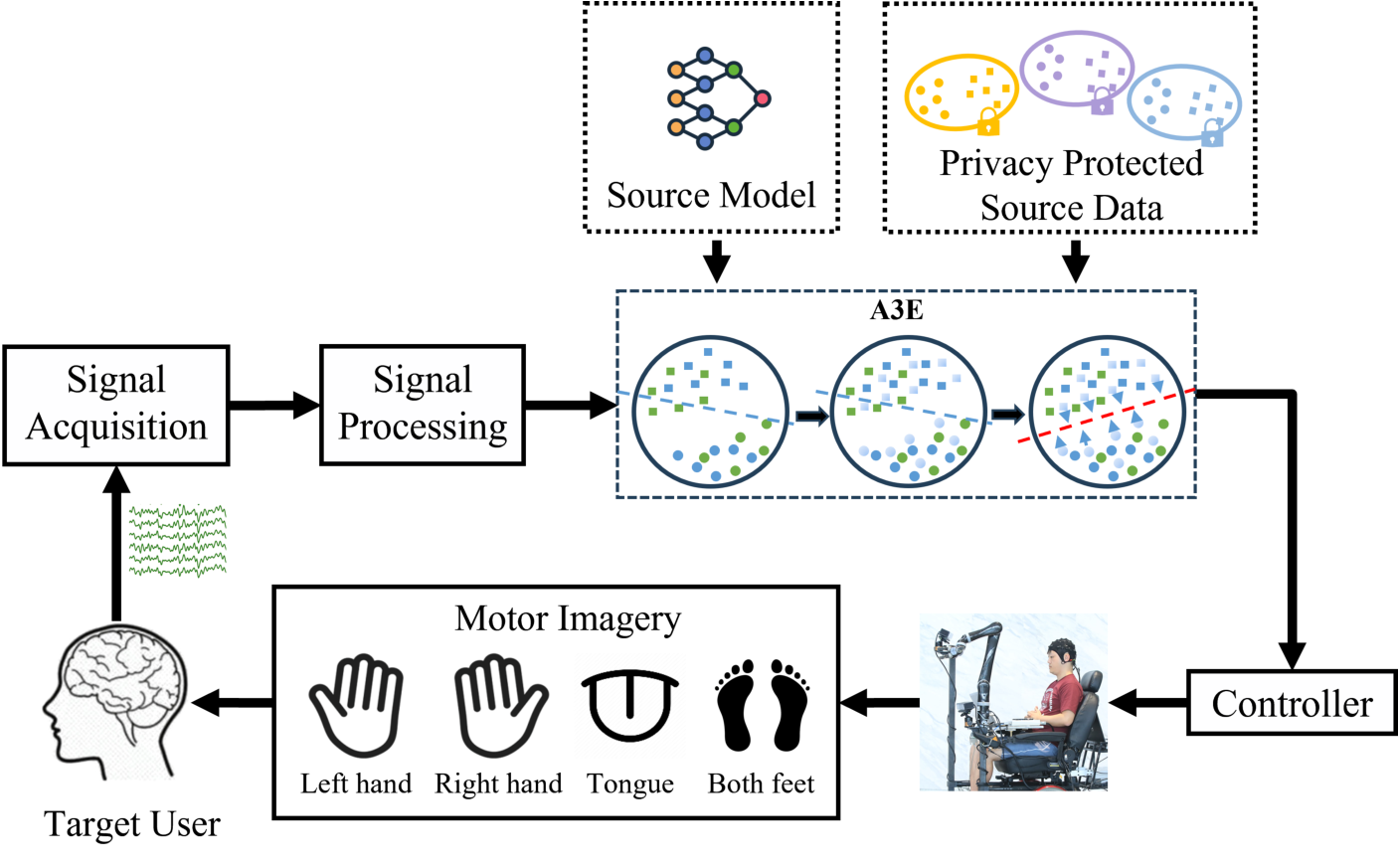}
\caption{The aligned and augmented adversarial ensemble (A3E) approach for simultaneously accurate decoding, adversarial robustness and privacy-protection in motor imagery BCIs.} \label{fig:A3E}
\end{figure}

\begin{enumerate}
\item \emph{Centralized source-free TL}, in which the source domains trust each other, so they can combine their EEG data together to train a single source model.
\item \emph{Federated source-free TL}, in which the source domains do not trust each other, so federated learning is used to isolate data from different source domains but still trains a single source model.
\item \emph{Source data perturbation}, in which a domain-specific perturbation is added to each domain to hide its private information (e.g., user identity), without hurting the motor imagery classification performance.
\end{enumerate}
The source model in the first two scenarios, or the perturbed and privacy protected source data in the last scenario, can then be provided to the target domain for TL. In this way, the source domain data privacy is protected. During TL, the A3E algorithm performs first EA and data augmentation in the target domain to increase the decoding accuracy on the benign samples, and then adversarial training to increase the robustness on the adversarial samples.

In this way, we can simultaneously achieve accurate decoding, adversarial robustness and privacy-protection in EEG-based BCIs. We expect EA will also play an important role in future such approaches.

\section{Conclusions} \label{sect:conclusions}

Due to the non-stationarity and large individual differences of EEG signals, EEG-based BCIs usually need subject-specific calibration to tailor the decoding algorithm for each new subject, which is time-consuming and user-unfriendly, hindering their real-world applications. TL has been extensively used to expedite the calibration, by making use of EEG data from other subjects/sessions. An important consideration in TL for EEG-based BCIs is to reduce the data distribution discrepancies among different subjects/session, to avoid negative transfer. EA was proposed in 2020 to address this challenge. Numerous experiments from 13 different BCI paradigms demonstrated its effectiveness and efficiency. This paper has revisited EA, explaining its procedure and correct usage, introducing its applications and extensions, and pointing out potential new research directions. It should be very helpful to BCI researchers, especially those who are working on EEG signal decoding.

Due to its effectiveness and efficiency, as suggested by Junqueira et al. \cite{Junqueira2024}, ``\emph{EA should be a standard pre-processing step when training cross-subject models.}"



\end{document}